\documentclass[twocolumn,showpacs,preprintnumbers,amsmath,amssymb,aps,pre]{revtex4}
\usepackage{epsfig}

\usepackage{graphicx,latexsym,xcolor}

\usepackage{tikz}

\usepackage[english]{babel}
\usepackage{amsmath}
\usepackage{color}
\usepackage{epsfig}
\usepackage{graphicx}
\usepackage{bm}
\usepackage{times,amsmath,amssymb}
\usepackage{subfigure}
\usepackage{amsfonts}
\usepackage{amssymb}
\usepackage{color}
\global\long\def\bege{\begin{equation}}
\global\long\def\ende{\end{equation}}

\global\long\def\begal{\begin{align}}
\global\long\def\endal{\end{align}}

\begin{document}

\title{Physics in non-fixed spatial dimensions via random networks}
\author{Ioannis Kleftogiannis$^1$, Ilias Amanatidis$^{2}$}
\affiliation{$^1$ Physics Division, National Center for Theoretical Sciences, Hsinchu 30013, Taiwan }
\affiliation{$^2$Department of Physics, Ben-Gurion University of the Negev, Beer-Sheva 84105, Israel}

\date{\today}
\begin{abstract}
We study the quantum statistical electronic properties of random networks which inherently lack a fixed spatial dimension. We use tools like the density of states (DOS) and the inverse participation ratio(IPR) to uncover various phenomena, such as unconventional properties of the energy spectrum and persistent localized states(PLS) at various energies, corresponding to quantum phases with  zero-dimensional(0D) and one-dimensional(1D) order. For small ratio of edges over vertices in the network $R$ we find properties resembling graphene/honeycomb lattices, like a similar DOS containing a linear dispersion relation at the band center at energy E=0. In addition we find PLS at various energies including E=-1,0,1 and others, for example related to the golden ratio. At E=0 the PLS lie at vertices that are not directly connected with an edge, due to partial bipartite symmetries of the random networks (0D order). At E=-1,1 the PLS lie mostly at pairs of vertices(bonds), while the rest of the PLS at other energies, like the ones related to the golden ratio, lie at lines of vertices of fixed length(1D order), at the spatial boundary of the network, resembling the edge states in confined graphene systems with zig-zag edges. As the ratio $R$ is increased the DOS of the network approaches the Wigner semi-circle, corresponding to random symmetric matrices(Hamiltonians) and the PLS are reduced and gradually disappear as the connectivity in the network increases. Finally we calculate the spatial dimension $D$ of the network and its fluctuations. We obtain both integer and non-integer $D$ and a logarithmic dependence on $R$. In addition we examine the relation of $D$ and its fluctuations to the electronic properties derived. Our results imply that universal physics can manifest in physical systems irrespectively of their spatial dimension. Relations to emergent spacetime in quantum and emergent gravity approaches are also discussed.
\end{abstract}

%\pacs{03.65.Nk, 05.60.-k, 05.40.Fb, 72.15.Rn, 73.63.Nm}
\maketitle

\section{Introduction}
The concept of spatial dimension has been central
in physics for many centuries. Physics problems are usually
formulated mathematically on geometrical manifolds with a
well defined spatial dimension, which is taken as an input. However in the last decades problems like quantum and emergent gravity and geometry, have shown that a reconsideration of this concept is required. For example models like, string theory formulated on continuous manifolds (membranes) of variable dimensions higher than three\cite{sunil,dienes}, or discrete/network/graph models like loop quantum gravity\cite{rovelli1,rovelli2},causal set cosmology\cite{bombelli,fay1,fay2,surya} the Wolfram-Gorard-Piskunov model\cite{wolfram,gorard} and others\cite{markopoulou,laughlin,lombard,verlinde,trugenberger,trugenberger2}, hint that spatial dimension as a fundamental concept has to be revised, if spacetime and its dimensionality are to be recovered as emergent from other more fundamental structures. Also, problems like the small value of the cosmological constant, which could be an instrisic property of spacetime as it emerges from other structures is a major related issue. All the above are strong suggestions that a theory reproducing the properties of spacetime in our universe, gravity and quantum effects could be spatio-dimensionless in its nature. Apart from quantum and emergent gravity, physics in non-fixed spatial dimensions can be useful in characterizing quantum and other phases emerging in many-body networks\cite{christandl,walschaers,ortega,many_body1}, that could be also relevant to emergent spacetime and its dimensionality, through entanglement.

In this paper we consider spatio-dimensionless physical systems, modelled via the most structureless models, random networks\cite{farkas,newman,frieze}, which are discrete mathematical models that do not inherently have a fixed spatial dimension. We consider uniform networks and study their quantum statistical electronic properties, using tools like the density of states(DOS) and the inverse participation ratio(IPR). We find that for small ratio of edges over vertices $R$ in the network, its electronic properties resemble those of graphene/honeycomb lattices with properties like a linear dispersion relation at the band center at energy E=0 and edge states\cite{fujita,fujita1,novoselov,geim,heiskanen} concentrating at the spatial boundary of the network. We find persistent localized states (PLS) at various energies including E=-1,0,1 and others, for example related to the golden ratio, via the study of the DOS and the IPR. These states comprise primarily of single unconnected vertices (0D order) at E=0 and one-dimensional(1D) clusters (1D order) for the other energies, where the wavefunction is localized. The energy of these 1D ordered states is determined by the dispersion of 1D tight-binding chains formed by the 1D clusters of various lengths. The 0D ordered states at E=0 spread over the whole network, while the 1D ordered states concentrate at the spatial boundary of the network, resembling the edge states in confined graphene systems with zig-zag edges\cite{fujita,fujita1,heiskanen}. As the number of edges between the vertices in the network increases, the DOS approaches the Wigner semicircle, corresponding to ensembles or random symmetric matrices, and the number of PLS is gradually reduced until they disappear for large $R$. Finally we calculate the spatial dimension D of the network. We find that the networks which resemble graphene have $D \approx 2$, fitting into a 2D plane. The dimension is increased logarithmically as $R$ increases and the network becomes more dense as more edges are added between its vertices. Additionally $D$ reaches integer values like $D \approx 4$ apart from non-integer ones. Our results show that universal phenomena can manifest in physical systems irrespectively of their spatial dimension. In addition we briefly discuss the relation of our results to emergent spacetime and relativity/gravity from discrete mathematical models.

\section{Network model}
Random networks are discrete mathematical models comprising
of many vertices randomly connected with edges\cite{farkas,newman,frieze}.
They can be used to express relations between different quantities, which is useful for example in simulating various real world behaviors such as propagation of behavioral patterns in social networks or in communications technology. Random networks are also useful in studying various localization phenomena in mesoscopic and random waveguide systems, theoretically and experimentally\cite{zhang}. The probability distribution of the number of edges at each vertex i (degree $d(i)$) determines the type of network. In the current paper we choose the most simple type, uniform networks (originally introduced by Paul Erdös and Alfred Rényi), whose degree at each vertex follows a hypergeometric function. Essentially we have a fixed number of vertices n and edges m, randomly distributed between them, forming a random network $G(n,m)$. All the configurations of the edges among the vertices, whose number is $\binom{ \binom{n}{2}}{m}$, have an equal probability to appear $p=\frac{1}{\binom{ \binom{n}{2}}{m}}$. The mean degree for each vertex in the uniform network is given by $<d(i)>=2\frac{m}{n}$ which can be interpreted as the average connectivity in the network. Note, that there are no multi-edges or loops namely the network is a simple undirected graph. We define the ratio of vertices over the edges in the network $R=\frac{m}{n}$, which is half its average connectivity $R=\frac{<d(i)>}{2}$.
We note that uniforms networks are a variant of 
random binomial graphs $G(n,p)$, where each edge is present with probability $p$. When $p=\frac{m}{\binom{n}{2}}$ uniform graphs and binomial graphs behave similarly in the limit $n\rightarrow\infty$ where $p \approx \frac{2m}{n^2}$.
For $R \le \frac{1}{2}$ the uniform network lies on the sub-critical phase, consisting of many small disconnected components. On the other hand, for $R>\frac{1}{2}$ the uniform network lies on the super-critical phase and contains instead a unique giant component and a few disconnected small components\cite{frieze}. In our calculations we consider $R>\frac{1}{2}$ ensuring that the network contains a large component to study.

We examine how a quantum particle behaves as it propagates through the tight-binding lattice formed by the random network. i.e. the electronic properties of the random network. The Hamiltonian of the system can be written as
\begin{equation}
H = \sum_{<i,j>}^m(c_{i}^{\dagger}c_{j} +  h.c.)
\label{ham}
\end{equation}
where $c_{i}^{\dagger}(c_{i})$ is the creation(annihilation) operation for a particle at vertex i in the random network. The indexes i,j are randomly sampled and create m pairs, representing the edges between the vertices in the network. The Eq. \ref{ham}, known also as the adjacency matrix of the network/graph, is a random matrix with a fixed number of elements m (the number of edges) with the value of one, randomly distributed inside it. In certain cases, such random network Hamiltonians have been shown to belong to the same universality class as models used to study Anderson localization phenomena\cite{zhang}. For example regular tight-binding lattices like the square or the cubic lattice with a random on-site potential (Anderson model). The randomly distributed elements in the Hamiltonian matrix Eq. \ref{ham} induce interference effects leading to localization of the wavefunctions as in the Anderson model or in random matrix theory(RMT) models. 
%%%%%%%%%%%%%%%%%
\begin{figure}
\begin{center}
\includegraphics[width=0.9\columnwidth,clip=true]{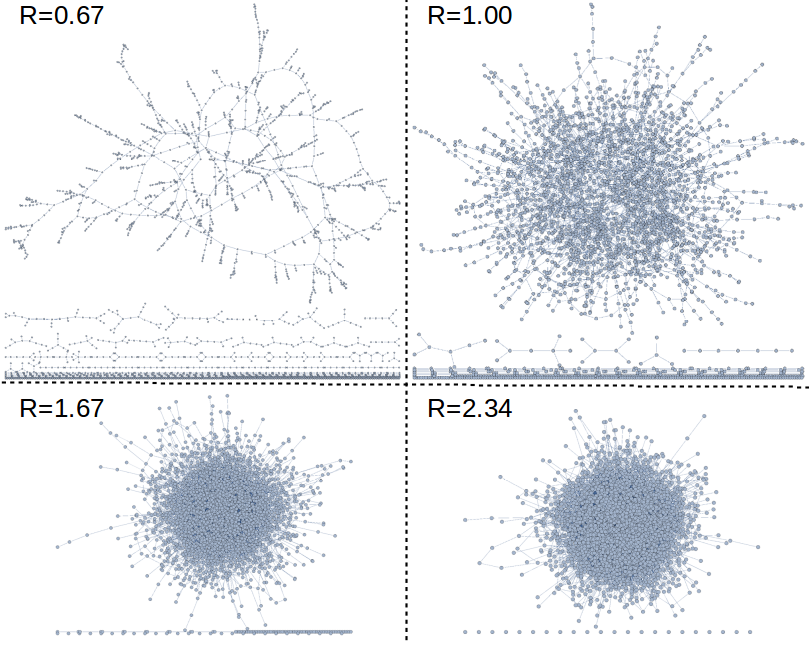}
\end{center}
\caption{Various random networks for different values of the ratio of the number of edges m over the number of vertices n ($R$). We fix n=3000 for all the cases shown. For $R=0.67$ the network is almost planar
with few overlapping/crossing edges, meaning that it can be fitted approximately into a 2D plane. As more edges are added $R$ increases and the network becomes spatially more dense.}
\label{fig1}
\end{figure}
%%%%%%%%%%%%%%%%%
\section{Energy spectrum}
\begin{figure}
\begin{center}
\includegraphics[width=0.9\columnwidth,clip=true]{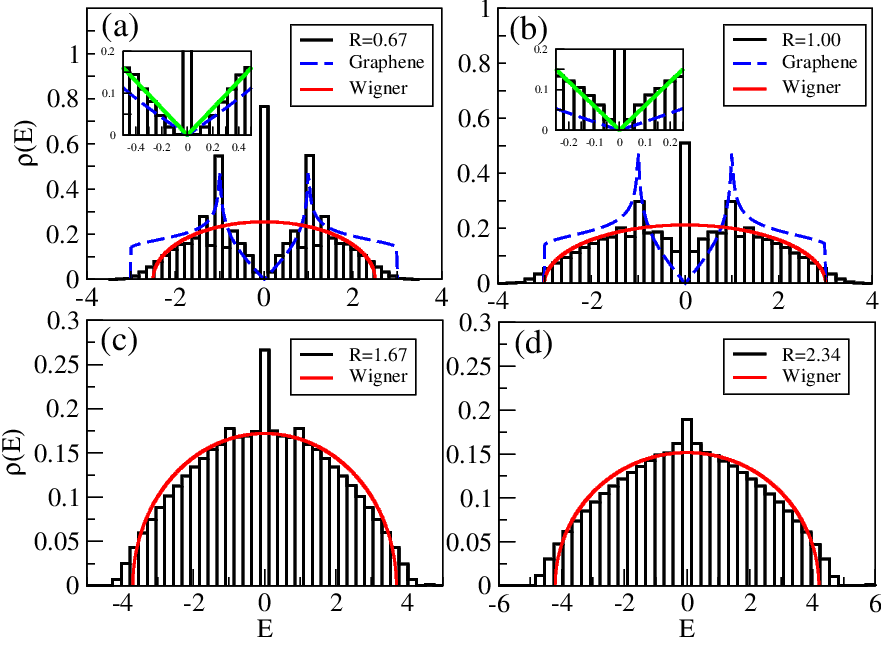}
\end{center}
\caption{The distribution of eigenvalues (density of states (DOS)) of the adjacency matrix for various values of the ratio of edges over vertices $R$ in the network, represented by the black histograms. We have considered 100 different configurations/realizations(runs) of the network. Also, isolated vertices have been removed from the calculation. The DOS in the upper panels (a) and (b) contain features from graphene/honeycomb lattice systems, as shown from the comparison with the DOS of graphene, represented by the blue dashed curve. A linear behavior of the DOS is shown in the insets, where the green solid curve is a linear fitting $\rho(E)=\alpha|E|$ with $\alpha=0.32$ for $R=0.67$ and $\alpha=0.6$ for $R=1$. As $R$ increases, the network becomes more dense and its DOS approaches the Wigner semi-circle in panels (c) and (d), represented by the red solid curves, resembling the distribution of eigenvalues of random symmetric matrices (Hamiltonians) from random matrix theory(RMT). A prominent peak at E=0 is present in all cases coming from many states at this energy, persistent for all networks, due to a partial bipartite symmetry of the network lattice.}
\label{fig2}
\end{figure}

In this section we examine the properties
of the energy spectrum of the random network
via the distribution of eigenvalues of Eq. \ref{ham}, i.e. the density of states (DOS) of the random network.
Firstly there are two limiting cases worth mentioning.
When all the n vertices in the network are disconnected (isolated) from each other (m=0) then all the elements of the Hamiltonian Eq. \ref{ham} are zero resulting in n zero eigenvalues and a DOS localized at energy E=0. On the other hand when all the vertices in the network are connected with each other, forming a k-complete graph, the Hamiltonian is full apart from its diagonal which has all its elements equal to zero. In the intermediate case, for an arbitrary number of vertices n and edges m in the network we have found two major forms for the DOS. The results are shown in Fig. \ref{fig2}. In the plots we have removed the eigenvalues E=0 coming from isolated vertices in the network which lead to lines/columns of zeros in the Hamiltonian.
%%%%%%%%%%%%%%%%%%%%%%%%%
%%%%%%%%%%%%%%%%%%%%%%%%%%%%%
For low ratio $R$ in Fig. \ref{fig2}(a),(b) the DOS of the random network ($\rho(E)$) resembles that of a graphene/honeycomb lattice as can be seen from the comparison between the blue dashed line and the histogram. The DOS of graphene is given by
\begin{equation}
\rho(E)=
\left\{ \begin{array}{ll}
\frac{1}{2\pi^{2}} \frac{4E}{(E+1)^{3} (3-E) } K\left( \sqrt{\frac{16}{(E+1)^{3} (3-E)}} \right) & 0\leq E < 1 \\
\frac{1}{2\pi^{2}} K\left ( \sqrt{\frac{(E+1)^{3} (3-E)}{16}} \right) & 1< E\leq 3 
\end{array} \right.
\end{equation}
for positive E, where K denotes the elliptic integral.
The DOS of graphene for negative E is produced by applying $E \rightarrow -E$ in the above equation. The main features of the DOS for both graphene and the network are two peaks at energies E=-1,1 and a linear behavior $\rho(E)=\alpha|E|$ near E=0 shown in the insets of Fig. \ref{fig2}(a),(b). The DOS of graphene reduces to the linear form $\rho(E) = \frac{1}{ 2 \pi \hbar^{2} v_{f}^{2}} |E|$ near E=0 where $v_{f} = \frac{\sqrt{3} a t}{2\hbar }$ is the Fermi velocity of the electrons and $a$ is the lattice constant. For $a=1$ and t=1 we have $\rho(E) = \frac{2}{ 3 \pi} |E|=0.21 |E|$ which is represented by the blue dashed curves in the insets of Fig. \ref{fig2}(a),(b). We note that for low $R$ there is a large probability that the network contains many small disconnected components as shown in Fig. \ref{fig1}. We have found that these components largely contribute in the number of states appearing at E=-1,1. Another main feature is a peak at E=0 which appears also for confined graphene systems with zig-zag edges like flakes and nanoribbons, due to edge states \cite{fujita,fujita1,heiskanen}. However the corresponding wavefunctions of the networks at E=0 are not edge states. States resembling the edge states in graphene, concentrating at the spatial boundary of the network, appear at energies away from E=0, as we shall show in the following sections where we examine their localization properties also. The resemblance to graphene can be seen schematically in the network presented in Fig. \ref{fig1} for $R=0.67$. The network consists primarily of polygons (cycles), whose corners consist of vertices that are connected mostly to three or four neighboring vertices with $<d(i)> \approx 3$. In addition there are very few crossing edges, i.e. the network is approximately planar, meaning that it can be fitted into a two-dimensional(2D) plane. This structure is topologically very similar to a graphene/honeycomb lattice which is a 2D plane of hexagons with every lattice site having three neighboring sites, i.e. the connectivity is three. The hexagons in graphene are topologically equivalent to the cycles in the network, irrespectively of their number of edges. Note that large uniform graphs, and virtually all random graphs, are locally tree-like. Cycles consisting of few edges, such as triangles, squares and pentagons are rare. When cycles appear they are usually very large, i.e., they consist of many edges. Therefore the local structure of graphs is very different from that of regular lattices, such as graphene where short cycles are abundant. Despite this fact, as we have shown from the analysis of the DOS, some of the electronic properties of sparse uniform networks resemble those of graphene. This is due to the similar lattice topologies of the two systems, i.e, the existence of cycles and the same connectivity between them ($<d(i)>=3$).

A linear dispersion relation $E \sim k$ can be derived from $\rho(E)=\alpha|E|$, by assuming a spherical symmetry in the Brillouin zone \cite{kollar}. Since the network for $R=0.67$ resembles a 2D manifold we can assume a coordinate system with two parameters x,y to describe it. We can define two wavevectors $k_{x}, k_{y}$ for a particle in a two-dimensional box
\begin{equation}
\label{schr_chiral}
\begin{array}{cc}
k_{x} =\frac{ \pi n_{x}  }{ L}
 \\
 \\
k_{y} =\frac{\pi  n_{y} }{ L}
\end{array}
\end{equation}
where L is the size of the box along the x and y directions.
%%%%%%%%%%%%%%%%%%%%%%%%%%%%%%%%%%%%%%%
The total number of states N in a volume/area inside the Brillouin zone is 
\begin{equation}
\label{schr_chiral}
N =\sum_{n_{x},n_{y} }\Delta n_{x}\Delta n_{y} =  \frac{L^2}{\pi^2}\sum_{k_{x},k_{y} }\Delta k_{x}\Delta k_{y}.
\end{equation}
%%%%%%%%%%%%%%%%%%%%%%%%%%%%%%%%%%%%%%%
Assuming a circular(rotational) symmetry in the Brillouin zone and using polar coordinates we can write
\begin{equation}
\label{schr_chiral}
\sum_{k_{x},k_{y} } \Delta k_{x}\Delta k_{y}= \int_{0}^{k} k dk \int_{0}^{2\pi} d \phi= \pi k^2,
\end{equation}
where $k^2= {k^2_{x}+k^2_{y}}$. Then we have
\begin{equation}
\label{schr_chiral}
N =\frac{ L^{2}k^{2}}{\pi}. 
\end{equation}
%%%%%%%%%%%%%%%%%%%%%%%%%%%%%%%%%%%%%%%
The total number of states per system area is
\begin{equation}
n=\frac{N}{L^{2}}= \frac{k^{2}}{\pi} => \\ 
\frac{dn}{dk} = \frac{2k}{\pi}.
\end{equation}
Then we can derive a linear energy dispersion relation E(k) by taking account of the linear behavior of the DOS near E=0 ($\rho(E)=\alpha|E|$) as follows,
\begin{equation}
\begin{split}
\rho(E) & =\frac{dn}{dE}=> \rho(E) dE =\frac{dn}{dk} dk => \\
\pi \int \alpha & |E| dE   = \int 2kdk =>  
E  =\sqrt{\frac{2}{\pi \alpha}} |k|.
\end{split}
\label{network_dispersion}
\end{equation}
The above dispersion is analogous to the linear energy dispersion for relativistic massless particles $E=\hbar ck$, followed also by graphene near E=0, where the speed of light c is replaced by the Fermi velocity of the electrons $v_{f} = \frac{\sqrt{3} a t}{2\hbar} $. For the networks from Eq. \ref{network_dispersion} we get an effective speed of light $v_{nt}=\frac{1}{\hbar}\sqrt{\frac{2}{\pi \alpha}}$, with dispersion $E=\hbar v_{nt}k$, where $\alpha$ can be obtained by the linearity(slope) of the DOS near E=0 from $\rho(E)=\alpha|E|$. For example we have $\alpha=0.32$ for $R=0.67$, $\alpha=0.6$ for $R=1$ and $\alpha=0.21$ for graphene. The velocity $v_{nt}$ could represent an upper limit for the transmission speed of information inside the network. Notice that a similar relation $v_{f}=\frac{1}{ \hbar \sqrt{2 \pi \alpha}}$ is followed also by graphene, meaning that $v_{nt}$ and $v_{f}$ will have the same order of magnitude ($10^6 m/s$), assuming that the hopping t and the lattice constant $a$ are the same for both graphene and the network.

The peaks at E=-1,0,1 and the linear dispersion near E=0, gradually disappear as $R$ increases as can be seen in Fig. \ref{fig2}(c),(d). As more edges are added in the network its Hamiltonian starts to resemble ensembles of random symmetric matrices, from RMT,  whose distribution of eigenvalues follows the Wigner semicircle, as the size of the matrices approach infinity.
This behavior has been observed in various uncorrelated random graphs\cite{farkas}.
The Wigner semicircle function is defined as
\begin{equation}
f(x)= \left\{
\begin{array}{ll}
    \frac{2}{\pi \textit{R}^2} \sqrt{\textit{R}^2- x^2}     & -\textit{R}\le x\le \textit{R} \\
     0 &  otherwise
\end{array}.
\right. 
\end{equation}
The above equation for various $\textit{R}$ is represented by the red solid curve in Fig. \ref{fig2} and describes sufficiently the DOS for the cases shown in the lower panels (c) and (d). Some features of the DOS at its edges are also captured by the Wigner semicircle in the upper panels (a) and (b) for low $R$. In overall we see that the DOS of the random network contains features from both graphene/honeycomb lattices and disordered systems described by RMT. The contribution of these two systems on the electronic properties of the network depends on the value of $R$, which determines the spatial density of the network. Sparse networks for low $R$ resemble graphene, while dense networks for large $R$ resemble disordered systems described by RMT.

The main difference with regular/periodic lattices such as the square, the cubic and the hypercubic is that their DOS reaches gradually a Gaussian distribution as the lattice connectivity, i.e. the number of neighbors at each site in the lattice goes to infinity. 

\section{Localization properties}
\begin{figure}
\begin{center}
\includegraphics[width=0.9\columnwidth,clip=true]{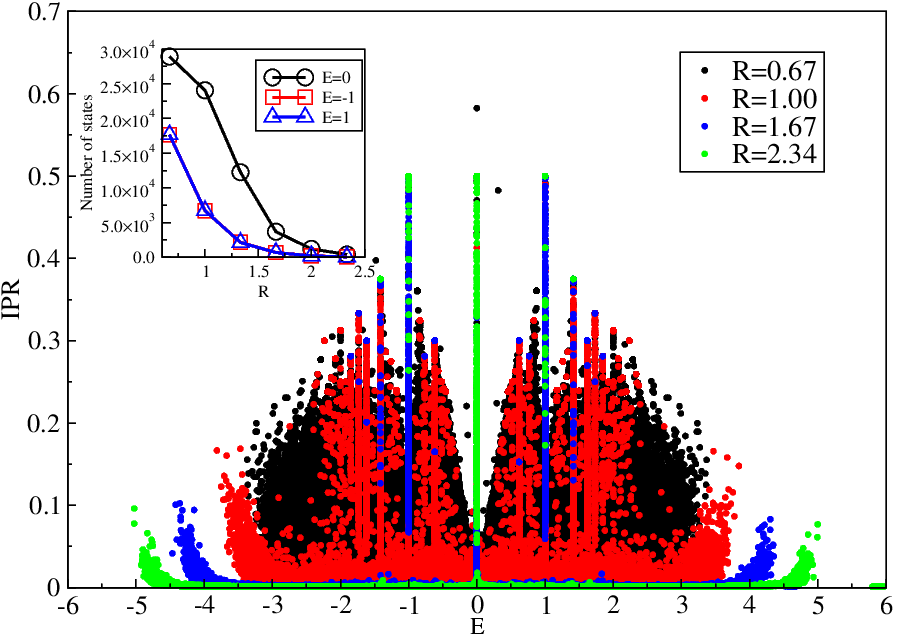}
\end{center}
\caption{The inverse participation ratio (IPR) for various values of the ratio $R$, for 100 runs. All isolated vertices have been removed from the calculation. The states across the whole energy spectrum become in average less localized as the network becomes more dense, since IPR decreases in average with increasing $R$. Persistent localized states (PLS) can be distinguished at various energies including E=-1,0,1
where IPR concentrates along vertical lines with higher values than the rest of the states.  As seen in the inset the number of PLS is gradually decreased with increasing $R$,
although it happens more slowly for the E=0 states.}
\label{fig3}
\end{figure}

In Fig. 3 we show the inverse participation ratio (IPR)
for the wavefunctions of Eq. \ref{ham} defined as
\begin{equation}
IPR(E) = \sum_{i=1}^n |\Psi_i(E)|^{4}
\label{ipr}
\end{equation}
where i runs over the all the vertices and $\Psi$ is the corresponding wavefunction amplitude. We show IPR for 100 different configurations/realizations of the network (runs). There are two major features that we can observe. As $R$ increases and the network becomes spatially more dense, IPR decreases in average. This means that the corresponding wavefunctions become in average less localized with increasing $R$. We have verified that these wavefunctions corresponding to low values of IPR spread over the whole network with random amplitudes at each vertex, resembling chaotic wavefunctions encountered in disordered tight-binding lattices\cite{ilias,paper0}. Another major feature that we can observe in Fig. \ref{fig3} is persistent localized states (PLS) appearing at various energies corresponding to high values of IPR, forming persistent vertical lines for all $R$. We can clearly distinguish PLS at energies E=-1,0,1 which correspond to peaks in the DOS shown in Fig. \ref{fig2}. As shown in the inset of Fig. \ref{fig3} the number of these states are reduced as $R$ increases. We have found that the states at E=0 are localized along the whole network, but only on vertices which are not directly connected with an edge. Note that these vertices should not confused with isolated vertices completely disconnected from the large component of the network, which have been removed from our calculations. All the E=0 states are localized on the same vertices albeit with different amplitudes. This is one main feature of the E=0 states for all $R$ and differs from confined graphene systems with zig-zag edges which contain edge states at E=0 concentrated at the boundary of the system \cite{fujita,fujita1,heiskanen}. We can consider that the PLS at E=0 in the network, form a quantum phase of zero-dimensional (0D) order, since the wavefunction lies only on vertices that are not directly connected with an edge. On the other hand we have found that the PLS at E=-1,1 localize mostly on pairs of vertices connected with an edge (bonds). These states concentrate at the spatial boundary of the network, i.e, they consist primarily of bonds connected to the periphery of the network with one vertex, resembling dangling bonds. This bond-ordered localized phase disappears for large $R$ indicating a phase transition at some critical value of $R$. On the other hand the E=0 states persist even for large $R$ as shown by the peaks in the DOS in Fig. \ref{fig2}. Other cases of PLS appear also at other energies for example at $E=-\phi,1-\phi,-1+\phi,\phi$ where $\phi=\frac{(1 + \sqrt{5})}{2}$ is the so called golden ratio. The wavefunctions of these states are localized primarily on lines of vertices (1D clusters) with fixed length, for example lines of four vertices for the energies related to the golden ratio, instead of bonds, and disappear for large $R$ as for the PLS at E=-1,1.
All the phases except those at E=0 can be considered as quantum phases of one-dimensional (1D) order. We have found that the energies of these 1D ordered states are coming from tight-binding chains whose length is determined by the length of the 1D clusters contained in the 1D ordered phases. The dispersion of a tight-binding chain with N sites and a hard-wall boundary condition at its ends is given by
\begin{equation}
E=2\cos{\frac{n \pi}{N+1}}, n=1,2,\dots,N.
\label{dispersion1d}
\end{equation}
The networks contain mostly eigenvalues of the above
equation for low N corresponding to short chains.
For example some of the E=-1,1 eigenvalues in the network come from Eq. \ref{dispersion1d} for N=2(chain with two sites) and the corresponding wavefunctions are localized along bonds of vertices. The case N=4 gives the energies related to the golden ratio. In general, the wavefunctions of the states in the network coming from energies described by Eq. \ref{dispersion1d}, apart from those at E=0, are localized along lines of vertices of length N and lie at the spatial boundary (periphery) of the network, connected to one vertex, resembling dangling bonds. In this sense these 1D ordered states at energies coming from Eq. \ref{dispersion1d} bear similarity to the edge states appearing in confined graphene systems, like flakes and ribbons, which are concentrated along the zig-zag edges of these systems\cite{fujita,fujita1,heiskanen} and persist even with disorder\cite{paper0}.The main difference with the networks is that the edge states in graphene appear at the Dirac point, at energy E=0, for example along flat bands for zig-zag nanoribbons \cite{fujita} or near E=0 for flakes, depending on their boundary morphology\cite{heiskanen}. Also, a large contribution to the eigenvalues Eq. \ref{dispersion1d} in the network comes from isolated chains of length N.
%%%%%%%%%%%%%%%%%%%%%%%%%%
%%%%%%%%%%%%%%%%%%%%%%%%%%

\section{Partial bipartite symmetry}
\begin{figure}
\begin{center}
\includegraphics[width=0.9\columnwidth,clip=true]{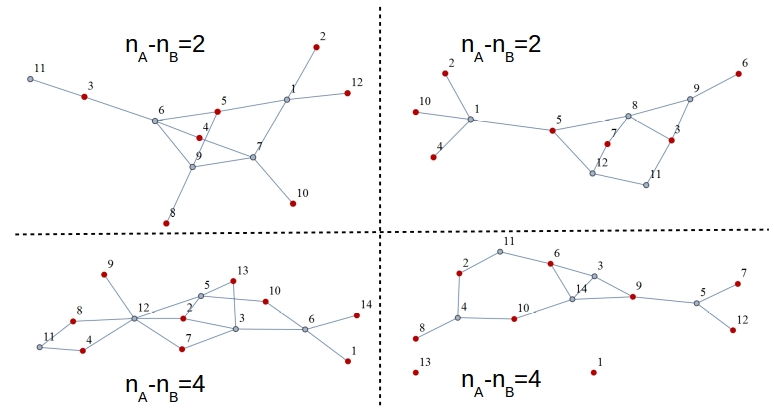}
\end{center}
\caption{Some examples of small networks demonstrating the partial bipartite symmetry. Red vertices form a sublattice of disconnected vertices (A), not directly connected with an edge, while the rest of the vertices form another sublattice (B). The difference in the number of vertices/sites between the two sublattices $n_{A} - n_{B}$ leads to an equal number of E=0 states. The wavefunction of these states is zero on the B sublattice which has the least number of vertices.}
\label{fig4}
\end{figure}

The appearance of the E=0 wavefunctions only on vertices
of the network that are not directly connected with
an edge can be explained as a consequence of a partial bipartite symmetry of the random networks\cite{inui,paper1,paper2}. The random network if seen as a lattice can always be split into two sublattices, say A and B with $n_A$ and $n_B$ sublattice sites respectively. Additionally it can always be splitted in such a way that there are no edges between the sites in one of the sublattices, say A. Some examples can be seen in Fig. \ref{fig4}. Then the Hamiltonian of the system can be simplified if written in the basis of A and B sites, as
%%%%%%%%%%%%%%%%%%%%%%%%%%%%%%
\begin{equation}
\label{ham_chiral}
H= \begin{bmatrix}
       0  & H_{AB} \\
    H_{AB}^{\dagger}    & H_{BB} 
\end{bmatrix}.
\end{equation}
We can write the Schr$\ddot{o}$dinger difference equations
centered on A and B sites as
%%%%%%%%%%%%%%%%%%%%%%%%%%%%%%
\begin{equation}
\label{schr_chiral}
\begin{array}{cc}

E\Psi_{A,i}= \sum_{j}\Psi_{B,j}
 \\

E\Psi_{B,i}= \sum_{j}\Psi_{A,j} + \sum_{j}\Psi_{B,j}.

\end{array}
\end{equation}
%%%%%%%%%%%%%%%%%%%%%%%%%%%%%%%
By setting E=0 we can see that the upper equation in Eq. \ref{schr_chiral} transforms to $\sum_{j}\Psi_{B,j}=0$ which is a set of $n_A$ equations with $n_B$ unknowns. If $n_A>n_B$ then  this set of equations can only be satisfied by setting $\Psi_{B,j}=0$ since there are more equations than unknowns i.e. the system of equations is overdetermined. Then the amplitudes on A sublattice $\Psi_{A,j}$ can be calculated by the lower equation in Eq. \ref{schr_chiral}, which is a system of $n_B$ equations with $n_A$ unknowns and gives $n_A-n_B$ linearly independent solutions. Consequently if the random network is split in two sublattices A and B, with one of them having no edges between its vertices, say A, whose number of vertices is larger than  the number of vertices for the other sublattice B, then there will always exist at least $n_A-n_B$ states at E=0. In addition the wavefunction of these states will have zero amplitudes on the sublattice with the smallest number of vertices, sublattice B. We remark that for periodic lattices like the square lattice which satisfy the full bipartite symmetry the value $n_A-n_B$ will depend on the shape of the boundary. For example for a square sample of square lattice we will always have $n_A-n_B=1$\cite{inui}, for odd side lengths, whereas in graphene flakes $n_A-n_B$ depends on the size of the system\cite{paper2}. Depending on the network configuration(run) there exist in general many different choices of the sublattice whose vertices are disconnected, corresponding to different partial bipartite symmetries. Note that in rare cases and more frequently in small networks with a low number of vertices and edges, some of equations for the sublattice A might be identical reducing the total number of equations for A. If the number of these equations becomes smaller than the number of B sites then the argument presented above breaks down and the wavefunctions spreads on both sublattices for the E=0 states. Nevertheless the partial bipartite symmetry is present for any type of random network and will lead to E=0 states with the properties described above in most cases of networks with low spatial density corresponding to low ratio of edges over vertices $R$.
%%%%%%%%%%%%%%%%%%%%%%%%%%%%%%%%%%
%%%%%%%%%%%%%%%%%%%%%%%%%%%%%%%%%%%%%5
\section{Spatial dimension}
%%%%%%%%%%%%%%%%%%%%%%%%%%
\begin{figure}
\begin{center}
\includegraphics[width=0.9\columnwidth,clip=true]{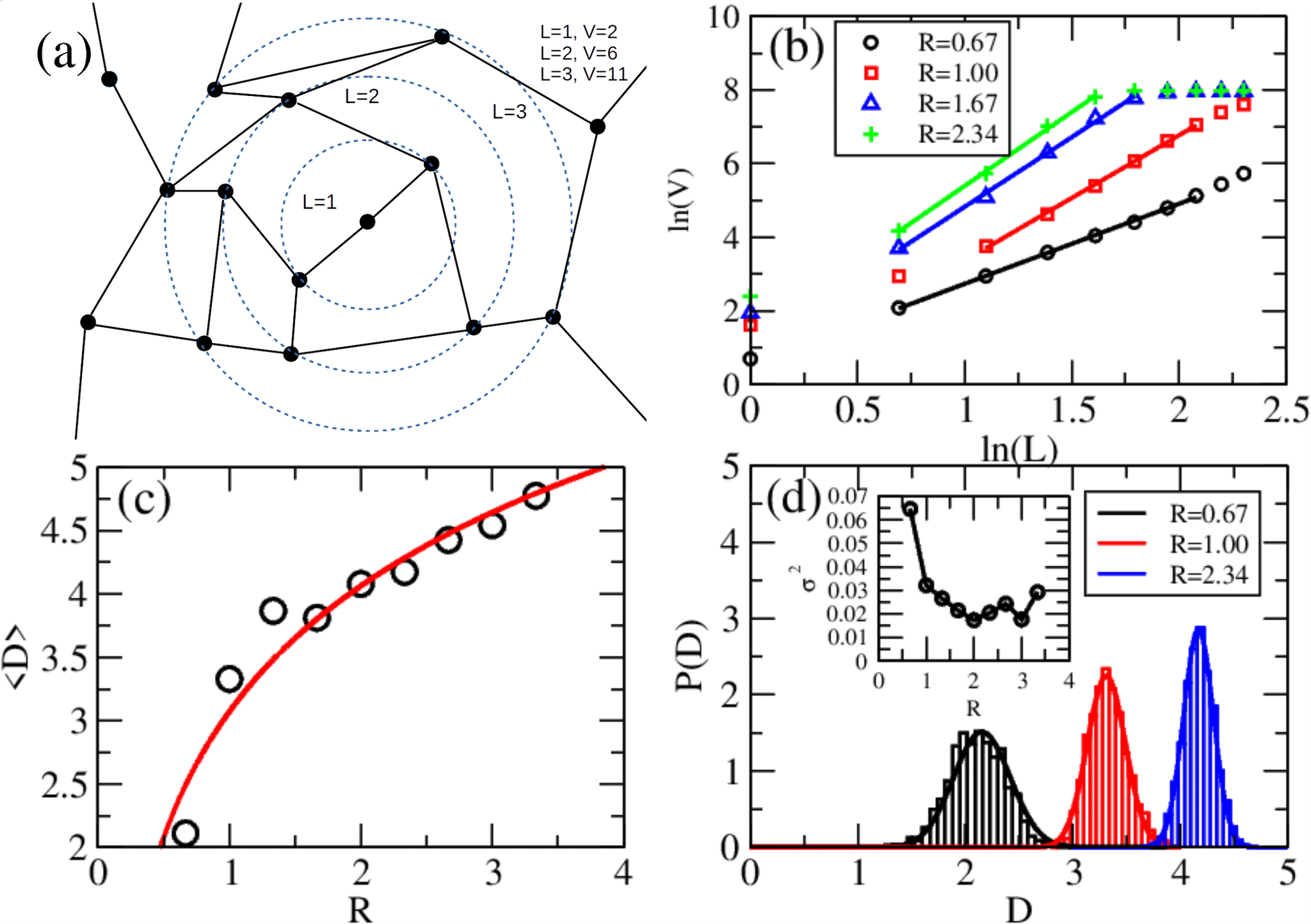}
\end{center}
\caption{(a) The process of calculating the spatial dimension D of the network by counting how the number of vertices contained into a topological sphere of radius L grows with L. The number of vertices added at each $L_{th}$ step, represented schematically by the circles, are determined by the edges/connections between the vertices (the central site for L=0 is not included in V). (b) The linear scaling of $ln(V)$ with $ln(L)$ gives the value of the dimension D from the relation $ln(V)=\beta+Dln(L)$ where $\beta$ is a fitting parameter. (c) The mean value of D (open circles) for 1000 configurations/realizations (runs) of the network versus the ratio $R$. Integer and non-integer values like $D \approx 4$ are reached. The dimension grows logarithmically with $R$ as $D = \gamma+\delta ln(R=\frac{<d(i)>}{2})$ represented by the red solid curve with fitting parameters $\gamma,\delta$. (d) The probability distribution of D follows the normal (Gaussian) distribution. As shown in the inset the variance $\sigma^2$ decreases with increasing $R$ and is minimized at $R=2$ where $D \approx 4$. }
\label{fig5}
\end{figure}
%%%%%%%%%%%%%%%%%%%%%%%%%%%
In this section we calculate the spatial dimension D of the random network, by examining how the connected vertices fill an ambient space. This dimension is related to the dimension of the ambient space in which the network can be embedded so that it contains no crossing edges. The values of D can be integer or non-integer. One example of this process can be seen schematically in Fig. \ref{fig5}(a). At each step in the scaling characterized by L we count the number of vertices that are connected directly with edges to the previous layer for L-1. This number is added in the overall number V, which counts the total number of vertices after the $L_{th}$ step, which can be thought also as the number of vertices contained in a topological sphere of radius L. Note that the circles in  Fig. \ref{fig5}(a) are for demonstration purpose only, with each circle simply denoting the $L_{th}$ step in the procedure of the calculation of D. Therefore by calculating the growth(scaling) of V with L we can calculate the dimension $D$ of the network via
\begin{equation}
V \sim L^D.
\label{dimension}
\end{equation}
We note that the above procedure bears similarity to the clustering growth method used to calculate the fractal/fractional dimension of an object which is already embedded in an ambient space with an integer topological dimension. This fractal dimension is not an integer in general and depends on the embedding space, i.e. on how the vertices have been arranged inside the coordinate system determined by the ambient space \cite{kosmidis}. This is unlike our case where the dimension is determined solely by the connections(edges) between the vertices without a predetermined embedding. Also, similar methods have been used in emergent space/geometry approaches \cite{wolfram,gorard,trugenberger2}. In the calculation of D we have considered only the largest component in the network, excluding the smaller disconnected ones.
Some results of this process for different $R$ can be
seen in Fig. \ref{fig5}(b). In addition in panel (c) we plot the average $<D>$ for 1000 runs (configurations/realizations of the random network). For $R=0.67$ the network is almost planar as $D \approx 2$, fitting approximately into a 2D plane without any crossing edges. This is one of the reasons that the sparse networks resemble graphene/honeycomb lattices and have similar electronic properties, as we have shown in the previous sections by the comparison between the DOS of the two systems. As the network becomes spatially more dense due to the increased connectivity, $D$ grows logarithmically with $R$ represented by the fitting red solid curve in Fig. \ref{fig5}(c). Non-Integer and integer dimensions like $D \approx 4$ are reached. However the electronic properties of the networks, as we have shown in the previous sections, do not resemble the respective regular lattices of integer dimension D, like the hypercubic lattices, whose DOS approaches a Gaussian distribution as $D$ increases. Instead the DOS of the random networks as $D$ increases approaches the Wigner semicircle, resembling disordered systems described by random Hamiltonians as in RMT. A linear dependence of the spatial dimension on the connectivity is followed by regular lattices like the chain, square, cube and hypercube. The connectivity $d(i)$ at each site in these lattices is two,four,six and eight giving the linear dependence $d(i)=2D \Rightarrow D= \frac{d(i)}{2}$. Since $R=\frac{<d(i)>}{2}$ our result in Fig. \ref{fig5}(c) shows that the dimension of the random network depends logarithmically on its average connectivity at each vertex, as $D \sim ln(<d(i)>)$, unlike the regular lattices. Also, this result implies an exponential dependence of the average connectivity in the network to D. Note that we expect large k-complete graphs to reach an infinite dimension, since every vertex will be connected to all other vertices. In Fig. \ref{fig5}(d) the probability distribution of D is shown along with its variance $\sigma^2$(fluctuations) in the inset. All the distributions can be fitted by normal (Gaussian) functions represented by the solid curves in Fig. \ref{fig5}(d). As can be seen in the inset, the dimensional fluctuations are reduced as $R$ increases, corresponding to higher dimension and spatially denser networks and are minimized at $R=2$ where $D \approx 4$.

\section{Emergent spacetime and relativity/gravity}
In this section we discuss briefly the relation of our model with the notion of emergent spacetime and relevant relativistic and gravitational effects. A natural way, in the sense of being the most structureless, lacking an inherent spatial dimension, would be to consider spacetime or space as a collection of random relations between abstract objects, sometimes called atoms of space\cite{rovelli1,rovelli2,bombelli,fay1,fay2,surya,wolfram,gorard,markopoulou,lombard,verlinde,trugenberger}, with various constraints. These relations can be naturally modeled by random networks, with the abstract objects(relations) represented by the vertices(edges), and the constraint in our case being the fixed ratio $R$ defined as the number of edges over the number of vertices. Essentially space could be represented as a quantum mechanical superposition of the different realizations of the random networks for fixed $R$. Then one of the main issues would be if the network can be reduced to a continuous manifold at a limit of large number of vertices and whether geometry can emerge in such a system. For instance, when $R=0.67$ in the random network that we studied, its spatial dimension is $D\approx2$ resembling a 2D plane. Moreover its electronic properties, as indicated by the DOS, resemble those of graphene, which is a 2D honeycomb lattice with the electrons following the Dirac equation near low energies E=0, at the Dirac points. At this energy, electrons behave relativistically as massless particles with the speed of light $c$ being replaced by the Fermi velocity of the electrons $v_f$, with a linear energy dispersion $E=\hbar v_{f} k$. The similarities of the electronic properties between graphene and spatially sparse random networks, pose an intriguing question, whether relativistic effects are also encountered in those random networks. As we have shown in the previous sections the effective speed of light in the networks is $v_{nt}=\frac{1}{\hbar}\sqrt{\frac{2}{\pi \alpha}}$ where $\alpha$ is determined by the DOS near E=0 from $\rho(E)=\alpha|E|$. The value of $v_{nt}$ could represent an upper limit for the transmission speed of information inside the network in the same way that c sets an upper limit for velocities in our universe.

Remarkably if we replace the parameters in the tight-binding lattice formed by the random network, with fundamental constants, for example the hopping energy t with the Planck energy $E_{p}$  ($t=E_p=\sqrt{\frac{\hbar c^5}{G}}$) and the lattice constant $a$ (edge length) with the Planck length $l_p$  ($a=l_p=\sqrt{\frac{\hbar G}{c^3}}$) then by taking account of the resemblance to graphene near E=0, we can estimate the order of magnitude of the network velocity, as $v_{nt} \sim \frac{at}{\hbar}=\frac{l_p E_p}{\hbar}=c$. Consequently by replacing the tight-binding parameters in the network with fundamental Planck constants, we get a network velocity that has the same order of magnitude as the speed of light ($10^8 m/s$).

We remark also that $D \approx 4$ is reached at $R=2$, when the number of edges is double the number of  vertices, where the dimensional fluctuations are minimized. This result would be tempting to relate to the 4D spacetime manifold in the models describing our universe, like in the Einstein field equations(EFE). In addition such a universe would be physically more stable, if its dimensional fluctuations were minimized, which we observe at $D \approx 4$. We note that time in network/graph models can be integrated either in the network itself\cite{fay1,fay2,surya} or can be treated as an external updating rule\cite{wolfram,gorard}. For example time in our model could be integrated as discrete evolution steps of updating the network, by adding an additional edge at each step. In this approach time could be represented by $R$ and would imply that the dimensionality of space could evolve with time.

In addition gravitational effects could potentially emerge in a random network approach to space. Various definitions of the curvature at each vertex in the network exist\cite{forman,dehoorn,ollivier,samal,knill,many-body2}, the simplest one being for tree and grid graphs
\begin{equation}
K(i)=1-\frac{d(i)}{2}.
\label{curvature}
\end{equation}
where $d(i)$ is the number of connected neighbors at vertex i in the network(degree). In the above definition triangles, tetrahedra and other higher n-dimensional cell complexes formed in the network structure are ignored. Since the average degree for a uniform network is $<d(i)>=2\frac{m}{n}$, using Eq. \ref{curvature}, we have $<K>=1-\frac{m}{n}$ for the curvature averaged over all the vertices in the network. We notice that $<K>$ transitions from a positive to negative value at $\frac{m}{n}=R=1$, where the dimension of the network is close to $D=3$ and $<K>=0$. This implies that if space is modeled with uniform random networks, having a zero or a slightly positive/negative curvature, which is the case for our current universe, would require a dimensionality close to $D=3$. Moreover the Ricci curvature tensor can be calculated in graph/network models\cite{forman,ollivier}. In addition geodesics can be defined in the network as simply the path between two vertices that has the least number of edges. Again an intriguing question arises, whether the curvature in the network can be related to the Riemannian curvature used in Einstein field equations(EFE) to describe how the geometry of spacetime, influenced by the mass-energy distribution, gives rise to gravitational effects.
Collections of vertices in the network could represent mass while edges can be related to the energy\cite{wolfram,gorard}. In this sense the PLS that we have found in the networks at various energies, comprising of lines of vertices (1D order), can be thought as emergent particles/excitations, whose mass is determined by the number of vertices in these 1D structures.

\begin{figure}
\begin{center}
\includegraphics[width=0.9\columnwidth,clip=true]{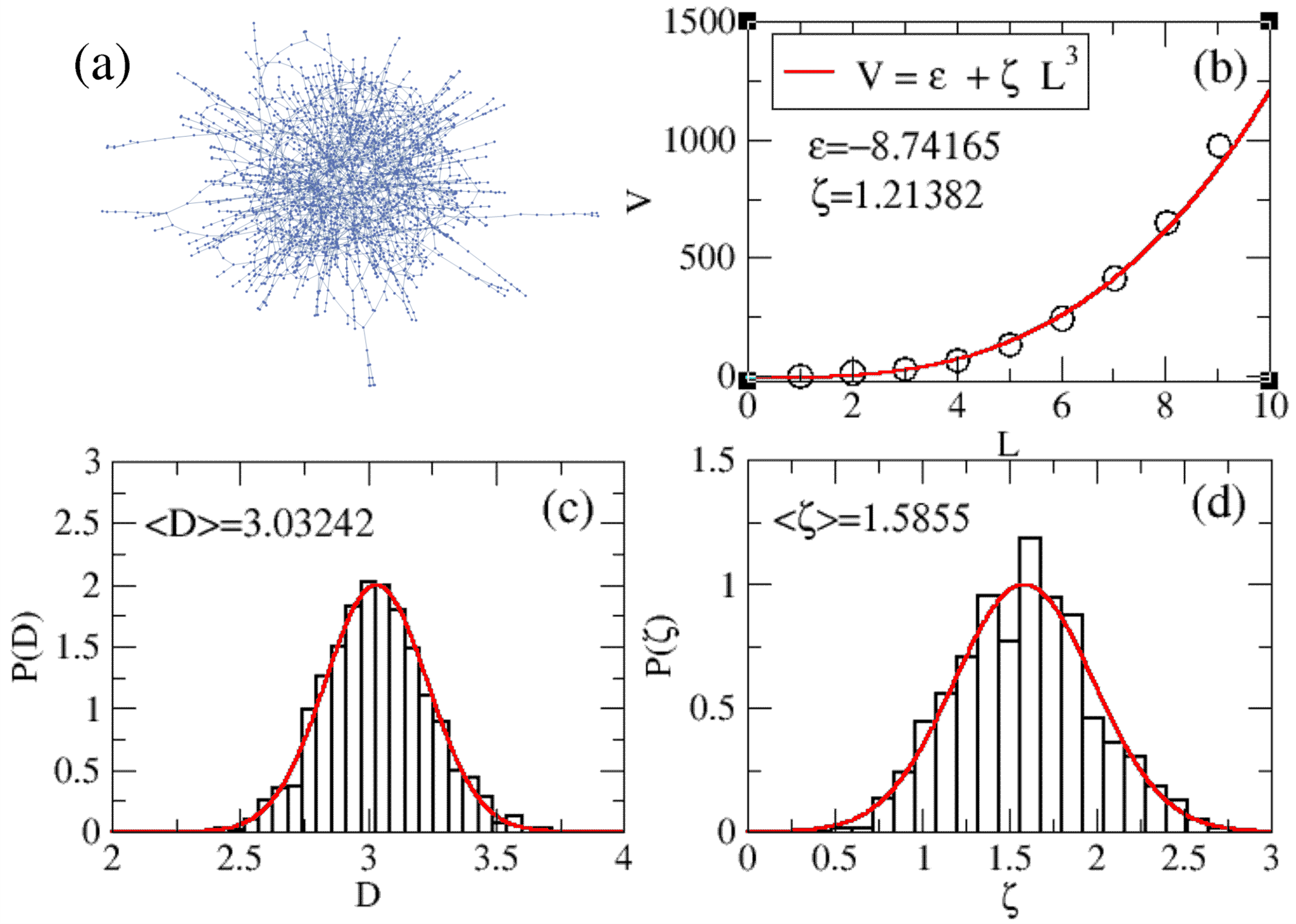}
\end{center}
\caption{(a) The largest component of the network with n=3000 and m=2750 corresponding to $R=0.92$. (b) The number of vertices scales as $V=\epsilon+\zeta L^{3}$ with L, with two fitting parameters $\epsilon$ and $\zeta$.(c)The fluctuations of the dimension D for the largest network component for 1000 runs. (d)The fluctuations of the fitting parameter $\zeta$.}
\label{fig6}
\end{figure}

Finally a random network approach to space wields a vertex density, which can be connected to the cosmological constant $\Lambda$ in EFE. The constant $\Lambda$, related to the energy density of the vacuum $\rho_{vacuum}=5.96 \times 10^{-27} \frac{kg}{m^{3}}$ in SI units, can be interpreted as an intrinsic property of space as it emerges from a random network model, like the uniform network that we have considered in the current paper. Since we have shown that the number of vertices V in the network grows approximately as $V=\epsilon+\zeta L^{D}$ then the density of vertices $\rho_{nt}=\frac{V}{L^D}$ is simply $\rho_{nt}=\epsilon L^{-D}+\zeta$. The constants $\epsilon$ and $\zeta$ can be related to $\Lambda$ or $\rho_{vacuum}$, for example near $R=1$ where $D \approx 3$, with the necessary substitutions of fundamental constants such as the Planck length $l_{p} = 1.616 \times 10^{-35} m$ in SI units. For example in Fig. \ref{fig6} we show the scaling of V vs L along with the fluctuations of D and $\zeta$ for a network with $R=0.92$ where $<D>=3.03242$. We have derived the value $<\zeta>=1.5855$ for this case. In order to make a comparison with the energy density of the vacuum we  consider its experimental value in kilograms(kg) per cubic meter($m^3$), $\rho_{vacuum}=5.96 \times 10^{-27} \frac{kg}{m^{3}}$. In addition we assume that the L in the scaling of the density of vertices $\rho_{nt}=\epsilon L^{-3}+\zeta$ is expressed in units of the Planck length $l_{p}=1.616 \times 10^{-35} m$. The first term $\epsilon L^{-3}$ can be ignored due to the large number of steps $L$ required to reach one cubic meter. Then by assigning a nominal mass $m_{v}$ to each vertex in the network, we get the total mass per cubic meter $\frac{\rho_{nt} m_{v}}{l^{3}_{p}}= \zeta m_{v} 0.237 \times 10^{105} \frac{1}{m^{3}}$. By taking account of $<\zeta>=1.5855$ for $R=0.92$, we can derive the value of $m_{v}$ by $<\zeta> m_{v} 0.237 \times 10^{105} \frac{1}{m^{3}} =\rho_{vacuum}= 5.96 \times 10^{-27} \frac{kg}{m^{3}} \Rightarrow m_{v}=1.59 \times 10^{-131} kg=8.88 \times 10^{-102} \frac{MeV}{c^{2}}$.
We note that the network model approach for space followed above contains essentially only two free parameters $\zeta$ and $m_{v}$ through the relation, 
\begin{equation}
\frac{\zeta m_{v}}{l^{3}_{p}}=\rho_{vacuum}.
\label{dimension}
\end{equation}
Particularly the value of $\zeta$ can be tuned so that $D=3$ for different types of random networks, other than uniform, that contain scaling relations like $V=\epsilon+\zeta L^{D}$.

The general question of how geometry arises from discrete mathematical models is a fundamental issue in both mathematics and physics. Another related problem is how to construct mathematics and physical theories in non-integer/fractional spatial dimensions or in the complete absence of them, like the approach that we follow in the current manuscript, by using random networks. 

\section{Summary and Conclusions}
We have studied the quantum statistical electronic properties of random networks which inherently lack a fixed spatial dimension. We have found that the electronic properties of the networks contain features from graphene/honeycomb lattice systems and disordered systems described by random Hamiltonians from random matrix theory (RMT). The similarities to graphene occur for spatially sparse networks with low ratio of edges over vertices $R$ and include features like a linear energy dispersion relation at the band center at energy E=0 and various persistent localized states (PLS). The PLS appear at various energies for example at E=-1,0,1 and others related for example to the golden ratio, and are localized either at single unconnected vertices (0D order) or along lines of vertices of fixed length which determines their energy (1D order).
The 1D ordered PLS concentrate at the spatial boundary(periphery) of the network, resembling the edge states in graphene systems with zig-zag edges.
As $R$ increases and the network becomes spatially more dense, its electronic properties, indicated for example by the DOS start to resemble disordered systems described by random Hamiltonians whose distribution of eigenvalues follows the Wigner semi-circle, as for uncorrelated random graphs. The PLS gradually disappear as $R$ increases, although those at E=0 persist even for large $R$. Finally we have calculated the spatial
dimension D of the network. We have found a logarithmic growth of D with $R$, reaching integer and non-integer values, implying an exponential dependence of the average connectivity in the network to D, unlike regular lattices. In summary we have studied quantum mechanics in physical systems lacking a fixed spatial dimension, demonstrating various unconventional electronic properties. These properties could be universal for quantum mechanical systems irrespectively of their spatial dimension. Finally we have discussed the relation of our results to spacetime and relativistic/gravity effects that could emerge from discrete mathematical models, like the random networks that we considered. In conclusion, we have offered an original approach that describes the emergence of different physical systems through the connectivity properties of quantum random networks.

\section*{Acknowledgements}
We acknowledge resources and financial support provided by the National Center for Theoretical Sciences of R.O.C. Taiwan and the Department of Physics of Ben-Gurion University of the Negev in Israel. Also, we acknowledge support by the Project HPC-EUROPA3 (INFRAIA-2016-1-730897), funded by the EC Research Innovation Action under the H2020 Programme. In particular, we gratefully acknowledge the computer resources and technical support provided by ARIS-GRNET and the hospitality of the Department of Physics at the University of Ioannina in Greece.

\end{document}